\documentclass[twocolumn,twoside,slac_two]{revtex4}
\usepackage{graphicx}
\usepackage{fancyhdr}
\input babarsym
\def\Btn{\ensuremath{B\ra\tau\nu}\xspace} 
\def\BDtn{\ensuremath{B\ra D^{(*)}\tau\nu}\xspace}
\def\EE{\ensuremath{E_{extra}}\xspace}

\def\tautoenunu {\ensuremath {\tau^+ \to e^+ \nu \nub}}

\def\tautomununu {\ensuremath {\tau^+ \to \mu^+ \nu \nub}}

\def\tautopinu {\ensuremath {\tau^+ \to \pi^+ \nub}}

\def\tautopipiznu {\ensuremath {\tau^+ \to \pi^+ \pi^{0} \nub}}

\begin{document}

\title{$B\ra\tau$ Leptonic and Semileptonic Decays}

%

\author{M. Barrett ---
On behalf of the \babar\ collaboration }
\affiliation{Brunel University, Uxbridge, United Kingdom}

\begin{abstract}
Decays of $B$ mesons to states involving $\tau$ leptons can be used as a  tool to search for the effects of new physics, such as those involving a charged Higgs boson.  The experimental status of the decays $B\ra\tau\nu$ and $\B\ra D^{(*)} \tau\nu$ is discussed, together with limits on new physics effects from current results. 
\end{abstract}

\maketitle

\thispagestyle{fancy}


\section{Introduction}
The decays of $B$ mesons to states containing a $\tau$ lepton offer the potential to look for the effects of new physics, such as the presence of a charged Higgs boson.  The effect of which would be greatly enhanced with respect to $B$ meson decays to lighter leptons, due to the Higgs coupling to mass.  $B\ra\tau X$ decays can be searched for in semileptonic, and fully leptonic decays of the $B$, and this presentation focuses on the searches for the decay modes $B\ra\tau\nu$, and $B\ra D^{(*)}\tau\nu$ (Charge conjugate modes are implied throughout).

\section{Recoil Analysis Technique}
\label{sec:recoil}
The search for $B$ meson decays involving a $\tau$ lepton is very challenging experimentally due to the presence of two or three neutrinos in the final state.  This leads to a lack of kinematic constraints on the signal $B$ meson ($B_{signal}$).  

All of the analyses discussed here use a recoil analysis technique in order to constrain $B_{signal}$.  This technique utilises the fact that pairs of B mesons are produced in the decays of $\Y4S$ particles, and thus finding the kinematical properties of one of the $B$ mesons will constrain the properties of the other $B$ meson.  Thus one $B$ meson in the decay (denoted $B_{reco}$ here) is fully reconstructed  --- this entails reconstructing all the decay products of $B_{reco}$ to establish its properties.  

There are two types of reconstruction used, which are referred to as ``Tags''.  The two types are called hadronic and semileptonic tags.  The general tags used by $\babar$ are given here.  Hadronic tags involve reconstructing the decay $B_{reco}\ra D^{*}X$, where $X$ is up to six light hadrons \{$\pipm, \piz, \Kpm, \KS$\}.  The $D^{*}$ mesons are reconstructed via $D^{*}\ra D\gamma$ and $D^{*}\ra D\piz$, and $D$ mesons are reconstructed $D^{0}\ra\Kp\pim$, $\Km\pip\piz$, $\Km\pip\pim\pip$, $\KS\pip\pim$ ($\KS\ra\pip\pim$).

Semileptonic tags involve reconstructing the decay $B\ra D\ell\nu X$, where $\ell$ is a light lepton \{$e$, $\mu$\}, and $X$ is $\piz, \gamma$, or nothing.  Fully reconstructed here assumes that one, and only one, neutrino is not reconstructed.  The $D$ mesons are reconstructed as above.    

On the signal side $\tau$ leptons may be reconstructed in up to 5 decay modes: $\tau\ra e \bar{\nu_{e}} \nu_{\tau}$, $\tau\ra\mu\bar{\nu_{\mu}}\nu_{\tau}$, $\tau\ra\pi\nu_{\tau}$, $\tau\ra\pi\piz\nu_{\tau}$, $\tau\ra\pi\pi\pi\nu_{\tau}$.  Together these modes make up approximately 80\% of the total $\tau$ lepton branching fraction.



\section{The Decay $B\ra\tau\nu$}
\subsection{Motivation}
The branching fraction for the decay $B\ra\ell\nu$ is given by:
\begin{equation}
\label{eqn:BRBtn}
\BR(B\ra\ell\nu) = \frac{G^{2}_{F} m_{B}}{8\pi} m^{2}_{l} \left( 1 - \frac{m^{2}_{l}}{m^{2}_{B}} \right)^{2} f^{2}_{B} |V_{ub}|^{2} \tau_{B},
\end{equation}
where $G_{F}$ is the Fermi constant, $m_{\ell}$ is the mass of the lepton, $f_{B}$ is the $B$ meson decay constant, $|V_{ub}|$ is an element of the CKM matrix, and $m_{B}$ and $\tau_{B}$ are the mass and mean lifetime, respectively, of the $B$ meson.

The lepton mass term in the equation leads to varying degrees of helicity supression between the different flavours --- the relative suppression is 1:$5\times10^{-3}$:$10^{-7}$ for $\tau$:$\mu$:$e$, leading to $B\ra\tau\nu$ having the highest branching fraction.
The $B$ meson decay constant can only be directly measured in purely leptonic $B$ meson decays, and theoretical predications from lattice QCD provide the best values for this quantity.  A current prediction of this value is $f_{B}=0.216\pm0.022\gev$ \cite{fB}.  

The CKM matrix element $V_{ub}$, describing the coupling between $b$ and $u$ flavoured quarks, has a current experimental value of $|V_{ub}|=(4.31\pm0.30)\times10^{-3}$ \cite{pdg}. 

\begin{figure}[h]
\begin{center}
\includegraphics[width=0.45\linewidth]{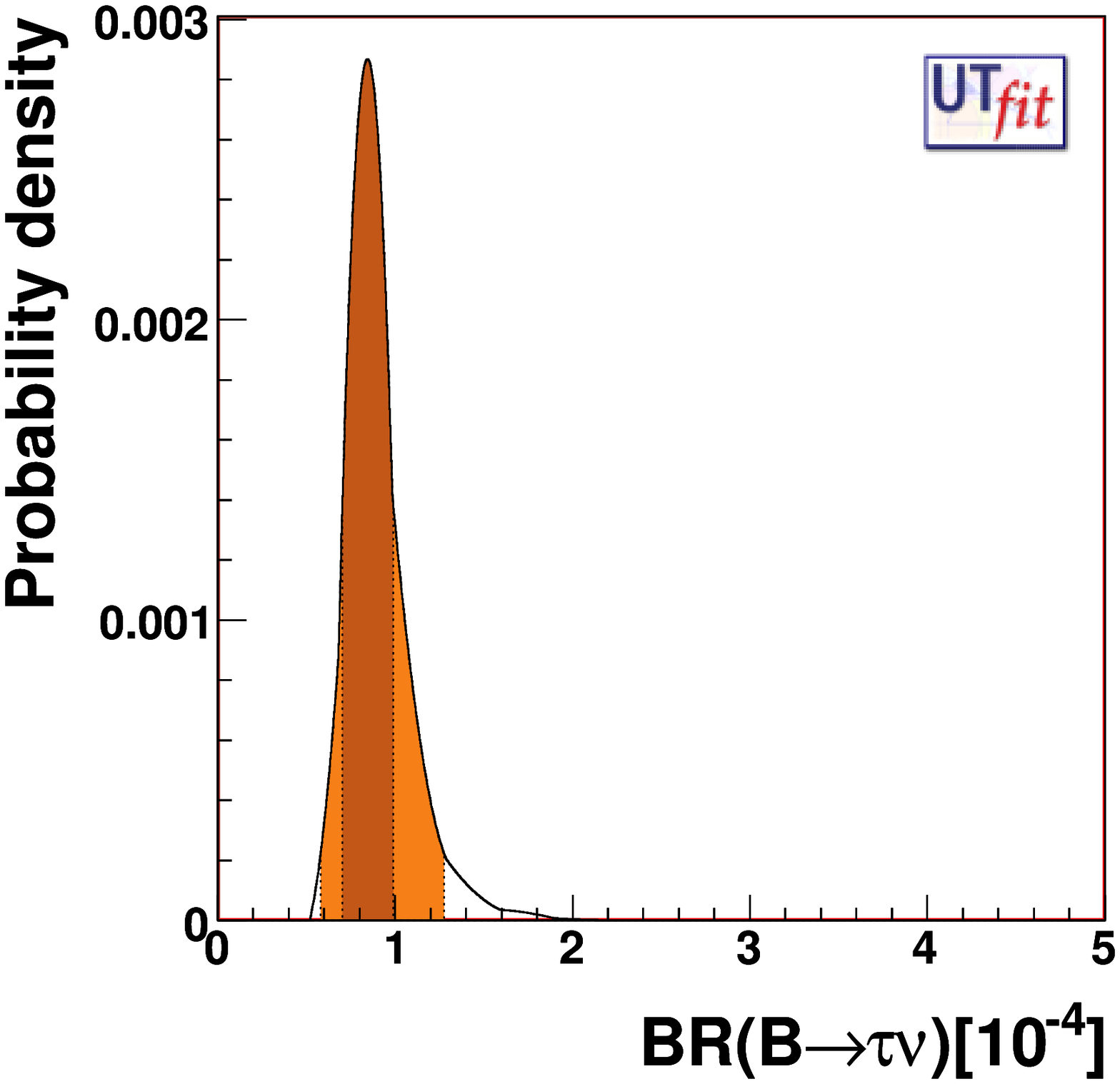}
\end{center}
\vspace{-0.5cm}
\caption{The predicted branching fraction for $\Btn$, from the UTfit collaboration \protect\cite{UTfit}, $\BR(B\ra\tau\nu) = (0.85\pm0.14)\times10^{-4}$, using $V_{ub}$, but no lattice calculations.}
\label{fig:UTfitPred}
\end{figure}

Taking the values above, gives a prediction for the Standard Model branching fraction of $\BR(B\ra\tau\nu) = (1.6\pm0.4)\times10^{-4}$.  The UTfit collaboration \cite{UTfit} produce a prediction for the value, shown in figure \ref{fig:UTfitPred}, which uses inputs from experimental results, including $V_{ub}$, but no predictions from lattice QCD, which gives a predicted value of $\BR(B\ra\tau\nu) = (0.85\pm0.14)\times10^{-4}$.

\begin{figure}[h]
\begin{center}
\includegraphics[width=0.45\linewidth]{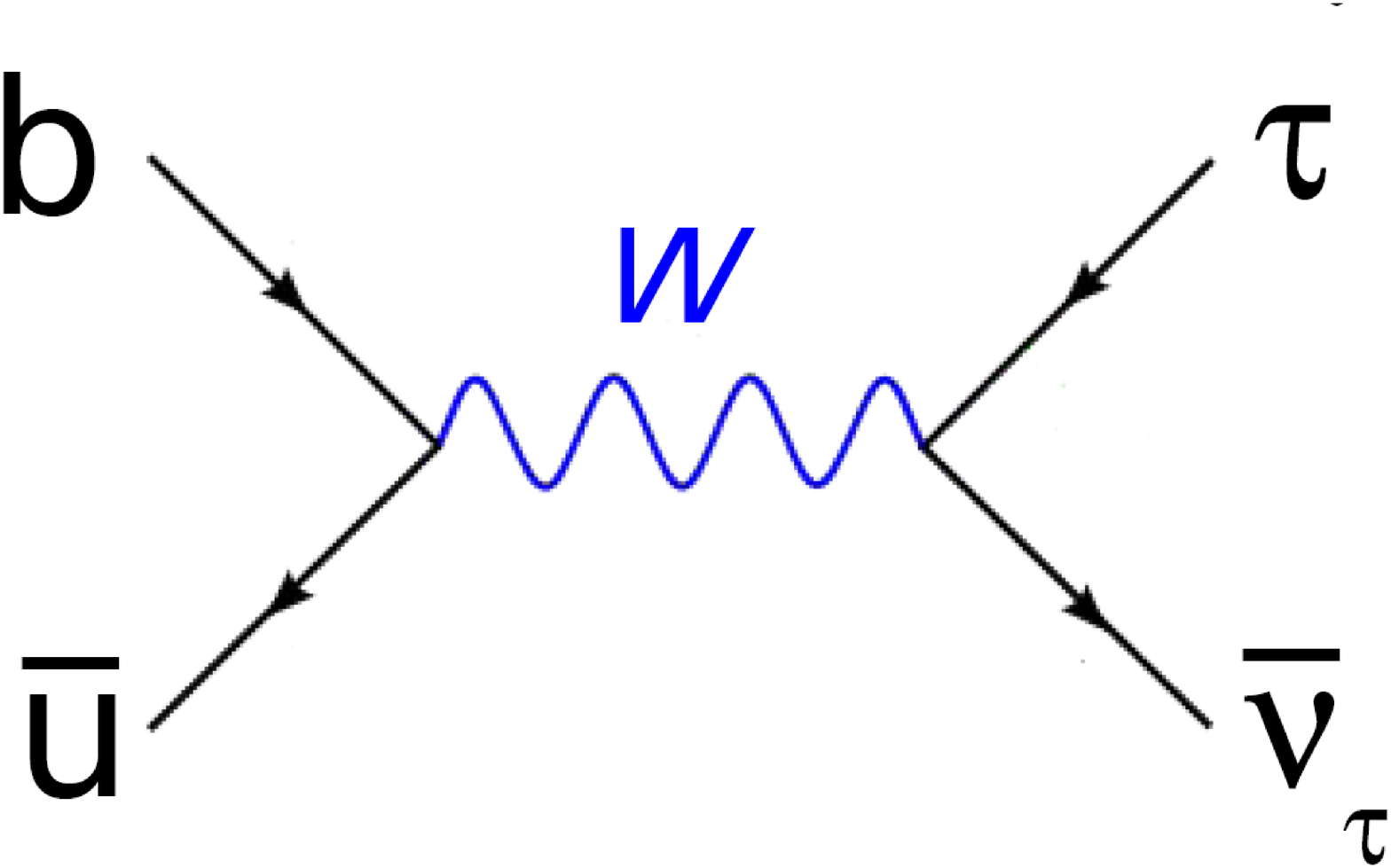}
\includegraphics[width=0.45\linewidth]{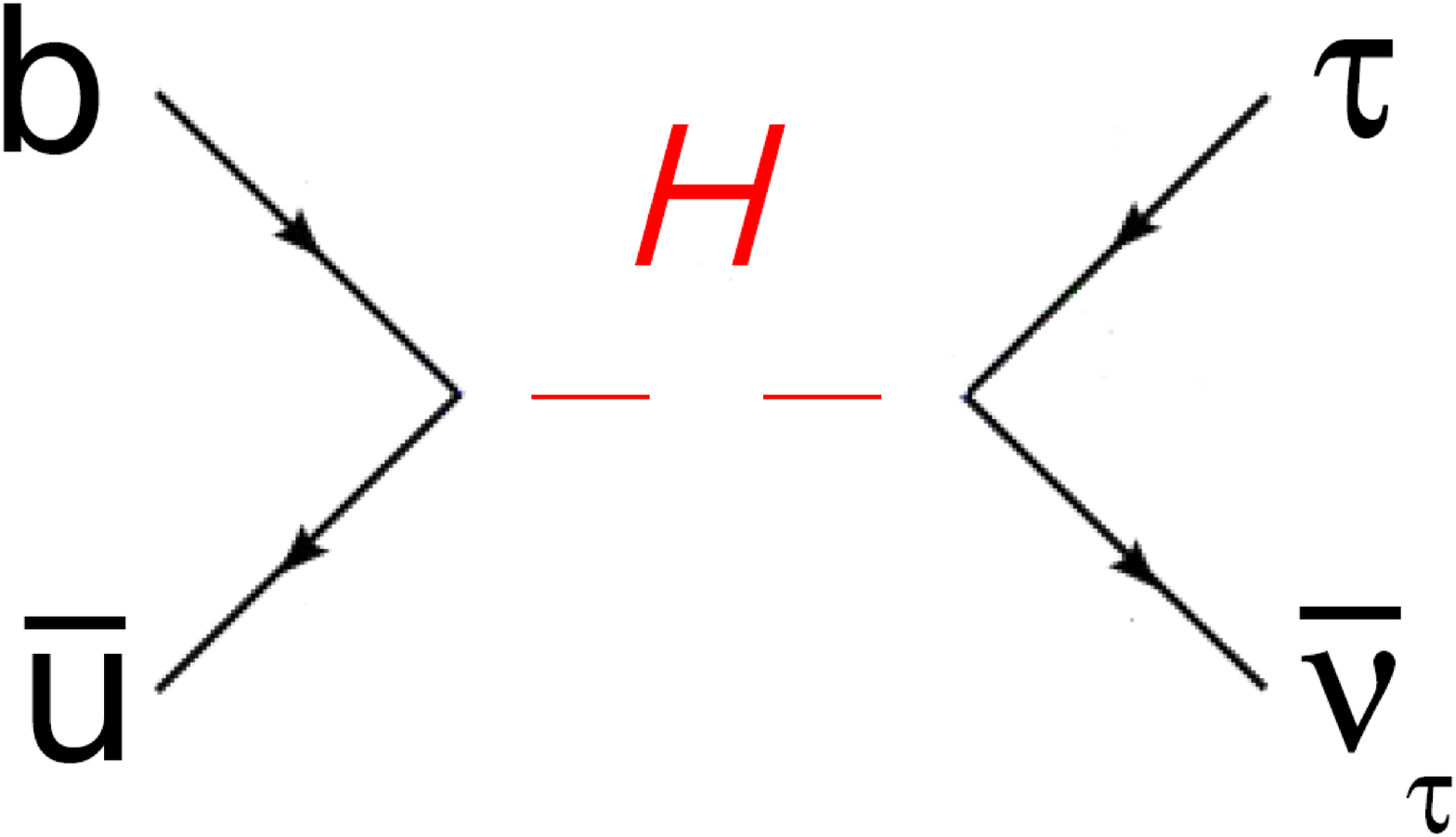}
\end{center}
\caption{Feynman diagrams showing the decay $\Btn$:  left --- Standard Model annihilation diagram; right --- extra diagram featuring a beyond the Standard Model charged Higgs boson.}
\label{fig:BtnFeyn}
\end{figure}

Beyond the Standard Model, an extra diagram is possible, replacing the $W$ boson in the annihilation diagram (figure \ref{fig:BtnFeyn}; left) with a charged Higgs boson (figure \ref{fig:BtnFeyn}; right).  In the Two Higgs Doublet Model (2HDM) \cite{WSHou} this modifies the branching fraction from its Standard Model value ($\BR^{SM}$):
\begin{equation}
\label{eqn:2HDM}
\BR^{2HDM} = \BR^{SM}\left(1-\frac{m^{2}_{B} \tan^{2} \beta}{m^{2}_{H^{\pm}}}\right)^{2},
\end{equation} 
where $m_{H^{\pm}}$ is the mass of the charged Higgs boson, and $\tan \beta$ is the ratio of the vacuum expectation values of the two Higgs doublets.

\subsection{Current Experimental Status}

\begin{figure}[htb]
\includegraphics[width=0.8\linewidth]{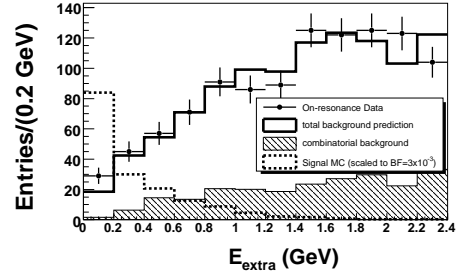} 
\caption{$\EE$ distribution after all selection criteria 
have been applied, for the \babar\ hadronic tags analysis \protect\cite{babarHad}. The on-resonance data (black dots) distribution is compared with the total background prediction
(continuous histogram). The hatched histogram represents the combinatorial background component.
$B^+\to\tau^+\nu$ signal MC (dashed histogram), normalized to a branching fraction of $3\times10^{-3}$ for illustrative purposes,
 is shown for comparison. }
\label{fig:babarHad}
\end{figure}

The \babar\ experiment has carried out two analyses of \Btn, using either hadronic and semileptonic tags.
The hadronic tag analysis \cite{babarHad} uses $383\times10^{6}$ \BB\ pairs. 
The variable \EE\ is used to define a signal region in the analysis, it is defined as the sum of all the energy in the calorimeter, that is not attributed to any of the reconstructed particles in the event.  For signal events it should peak at, or near, 0$\gev$.  The distribution of \EE\ is shown in figure \ref{fig:babarHad} -- the signal Monte Carlo (MC) distribution is plotted for a $\Btn$ branching fraction of $10^{-3}$ to allow for comparison of the shapes of the distributions.

The analysis is carried out using four decay modes of the $\tau$ lepton.  For each of these modes a signal region is defined, by selecting events with $\EE \lesssim 0.3\gev$ (the actual value varies slightly between modes).  In each mode the number of expected background events is compared with the number of observed events.  The yields in the four decay channels can be seen in table \ref{tab:babarHad}. 

\begin{table}
\caption{
Observed number of on-resonance data events in the signal region compared
 with the number of expected background events, for the \babar\ analysis using hadronic tags \protect\cite{babarHad}.}
\begin{center}
\begin{tabular}{lcc} \hline 
$\tau$ decay mode   & Expected background  &  Observed   \\ 
\hline \hline
\tautoenunu	  & 1.47  $\pm$ 1.37   & 4  \\ 
\tautomununu      & 1.78  $\pm$ 0.97   & 5  \\ 
\tautopinu        & 6.79  $\pm$ 2.11   & 10 \\ 
\tautopipiznu     & 4.23  $\pm$ 1.39   & 5  \\ 
\hline
All modes    & 14.27 $\pm$  3.03 & 24  \\ \hline \hline
\end{tabular}
\end{center}
\label{tab:babarHad}
\end{table}

The branching fraction is extracted by carrying out a likelihood fit to the yields in the four channels, this gives a branching fraction of $\BR(\Btn) = (1.8^{+0.9}_{-0.8}\pm0.4\pm0.2)\times10^{-4}$, where the first error is statistical, the second is due to background uncertainty, and the third is systematic.  An upper limit is placed at $\BR(\Btn) < 3.4\times10^{-4}$.  The product of $f_{B}$ and $V_{ub}$ is extracted as $f_{B} \cdot |V_{ub}| = (10.1^{+2.3}_{-2.5} (\mathrm{stat.}) ^{+1.2}_{-1.5} (\mathrm{syst.}))\times10^{-4}\gev$.      

\begin{figure}[h]
\begin{center}
\includegraphics[width=0.8\linewidth]{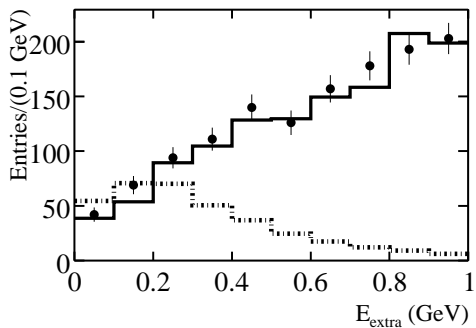}
\end{center}
\vspace{-0.8cm}
\caption{\EE\ distribution after all selection criteria 
have been applied, for the \babar\ semileptonic analysis. 
Background MC (solid histogram) has been normalized to the luminosity 
of the on-resonance data (black dots), and then
additionally scaled according
to the ratio of predicted background.
$\Btn$ signal MC (dotted histogram) is normalized to a branching
fraction of $10^{-3}$ and shown for comparison.}
\label{fig:babarSL}
\end{figure}

The semileptonic tagged analysis performed at \babar\ also uses $383\times10^{6}$ \BB\ pairs.  It uses a similar analysis strategy to extract the branching fraction from the observed yield in four channels.  The expected and observed yields are shown in table \ref{tab:babarSL}, and the \EE\ distribution in figure \ref{fig:babarSL}.  

\begin{table}[hbt]
\centering
\caption{\label{tab:babarSL}
Observed number of data events in the signal region, 
together with number of expected background events, for the \babar\ analysis using leptonic tags \protect\cite{babarSL}.}
\begin{tabular}{lcc} \hline \hline
$\tau$ decay mode & Expected background  &  Observed events  \\ 
	          & events               &  in on-resonance data  \\ \hline
$\tautoenunu$     & 44.3  $\pm$ 5.2   & 59  \\ 
$\tautomununu$    & 39.8  $\pm$ 4.4   & 43  \\ 
$\tautopinu$      & 120.3 $\pm$ 10.2  & 125  \\ 
$\tautopipiznu$   & 17.3  $\pm$ 3.3   & 18  \\ 
\hline
All modes    & 221.7 $\pm$ 12.7  & 245  \\ \hline \hline
\end{tabular}
\end{table}

The extracted branching fraction is $\BR(\Btn) = (0.9\pm0.6\pm0.1)\times10^{-4}$, where the first error is statistical, and the second systematic.  The upper limit derived from this is: $\BR(\Btn) < 1.7\times10^{-4}$.  The product $f_{B} \cdot |V_{ub}| = (7.2^{+2.0}_{-2.8}(\mathrm{stat.})\pm0.2(\mathrm{syst.})\times10^{-4}$ is also calculated.

The two tags used by \babar\ are statistically independent, and can be combined together, by extending the likelihood fit to encompass all eight channels.  Doing so yields the result $\BR(\Btn) = (1.2\pm0.4(\mathrm{stat.})\pm0.3(\mathrm{bkg.})\pm0.2(\mathrm{syst.}))\times10^{-4}$, which represents a statistical significance of $2.6\sigma$ (standard deviations).  

\begin{figure}[h]
\begin{center}
\includegraphics[width=0.45\linewidth]{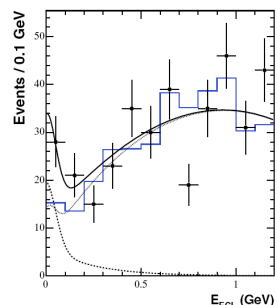}
\end{center}
\vspace{-0.8cm}
\caption{Distributions of $E_{ECL}$ from the Belle analysis \protect\cite{belleBtn} of $\Btn$.  The data are shown as points, and background MC as a solid histogram.  The solid curve shows the result of a fit, made of signal (dashed curve), and background (dotted curve) contributions.}
\label{fig:belleBtn}
\end{figure}

 The Belle analysis of \Btn\ uses $449\times10^{6}$ \BB\ pairs.  It uses hadronic tags, reconstructing $B\ra D^{(*)0}X$, where $X$ is $\pi$, $\rho$, $a_{1}$, or $D_{s}$.  Five $\tau$ lepton decay modes are used to determine the overall branching fraction.  The distribution of the extra calorimeter energy, denoted $E_{ECL}$, is shown in figure \ref{fig:belleBtn}.
 The obtained result is $\BR(\Btn) = (1.79^{+0.56}_{-0.49}(\mathrm{stat.}^{+0.46}_{-0.51}(\mathrm{syst.}))\times10^{-4}$, representing a statistical significance of 3.5$\sigma$.  The quantity $f_{B} \cdot |V_{ub}| = (10.1^{+1.6}_{-1.4}(\mathrm{stat.})^{+1.3}_{-1.4}(\mathrm{syst.}))\times10^{-4}\gev$ is also reported.

\begin{figure}[h]
\begin{center}
\includegraphics[width=0.45\linewidth]{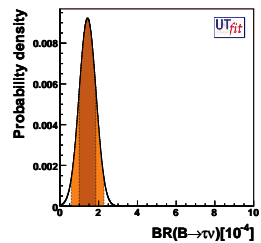}
\end{center}
\vspace{-0.5cm}
\caption{The combined branching fraction $\BR(\Btn)= (1.41\pm0.43)\times10^{-4}$, based on experimental measurements of this mode, from the UTfit collaboration \protect\cite{UTfit}.}
\label{fig:UTfitExp}
\end{figure}

Figure \ref{fig:UTfitExp} shows a plot from the UTfit collaboration which combines the experimental results for $\BR(\Btn)$, from which they extract a central value of $\BR(\Btn)= (1.41\pm0.43)\times10^{-4}$.

The experimental results from $\Btn$ can be used to constrain physical effects within and beyond the Standard Model.  The branching fraction given in equation \ref{eqn:BRBtn}, can be combined with the neutral $B$ meson oscillation parameter:

\begin{equation}
\Delta m_{d} = \frac{G^{2}_{F}}{6\pi^{2}}\eta_{B}m_{B}m^{2}_{W}f^{2}_{B}B_{B}S_{0}(x_{t})|V_{td}|^{2},
\end{equation}
where $G_{F}$ is the Fermi constant, $\eta_{B}$ is a QCD correction factor (dependent on $\Lambda_{QCD}$, and the quark masses $m_{b}$, and $m_{t}$), $m_{B}$ is the mass of a $B$ meson, $m_{W}$ is the mass of a W $boson$, $f_{B}$ is the $B$ meson decay constant, $B_{B}$ is the bag parameter arising from the vacuum insertion approximation, $S_{0}(x_{t})$ is the Inami-Lim function, and $|V_{td}|$ is an element of the CKM matrix.

\begin{figure}[h]
\begin{center}
\includegraphics[width=0.6\linewidth]{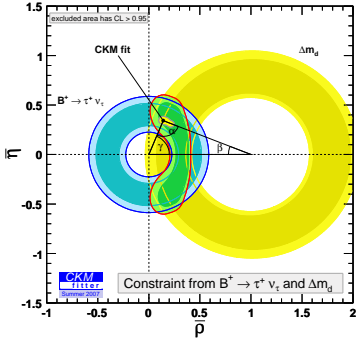}
\end{center}
\vspace{-0.5cm}
\caption{Constraints on the Unitarity Triangle, from the measurements of $\BR(\Btn)$ (smaller blue annulus), and $\Delta m_{d}$ (larger yellow annulus).  Produced by the CKMfitter collaboration \protect\cite{CKMfitter}.} 
\label{fig:CKMUT}
\end{figure}

By taking the ratio of these two quantities, the parameter $f_{B}$, which is the least well determined, cancels out, and the ratio of the CKM matrix elements $|V_{ub}| / |V_{td}|$ can be extracted.  This can be represented graphically as a constraint on the apex of the Unitarity Triangle, and figure \ref{fig:CKMUT}, produced by the the CKMfitter \cite{CKMfitter} collaboration shows this.

\begin{figure}[h]
\begin{center}
\includegraphics[width=0.45\linewidth]{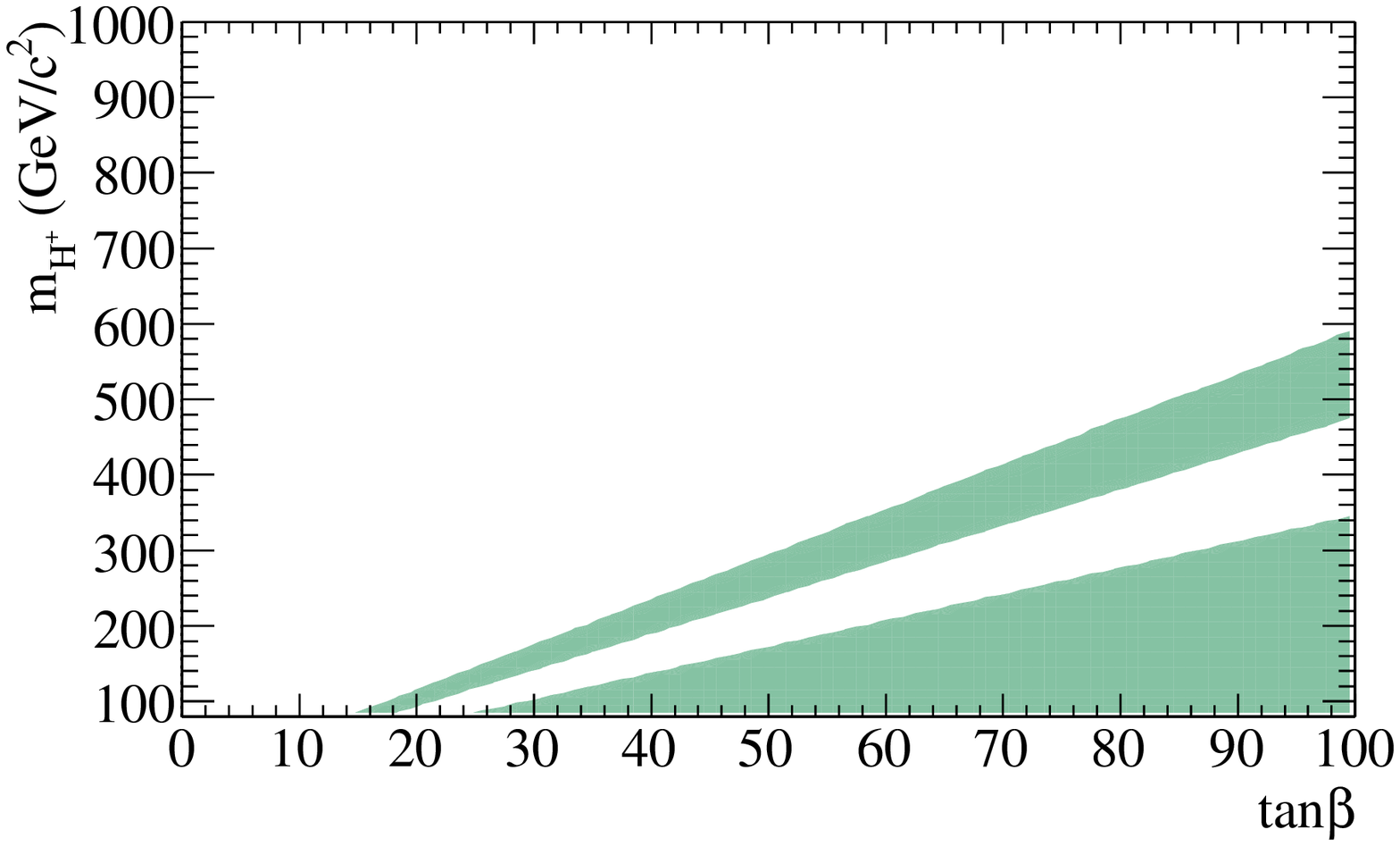}
\includegraphics[width=0.45\linewidth]{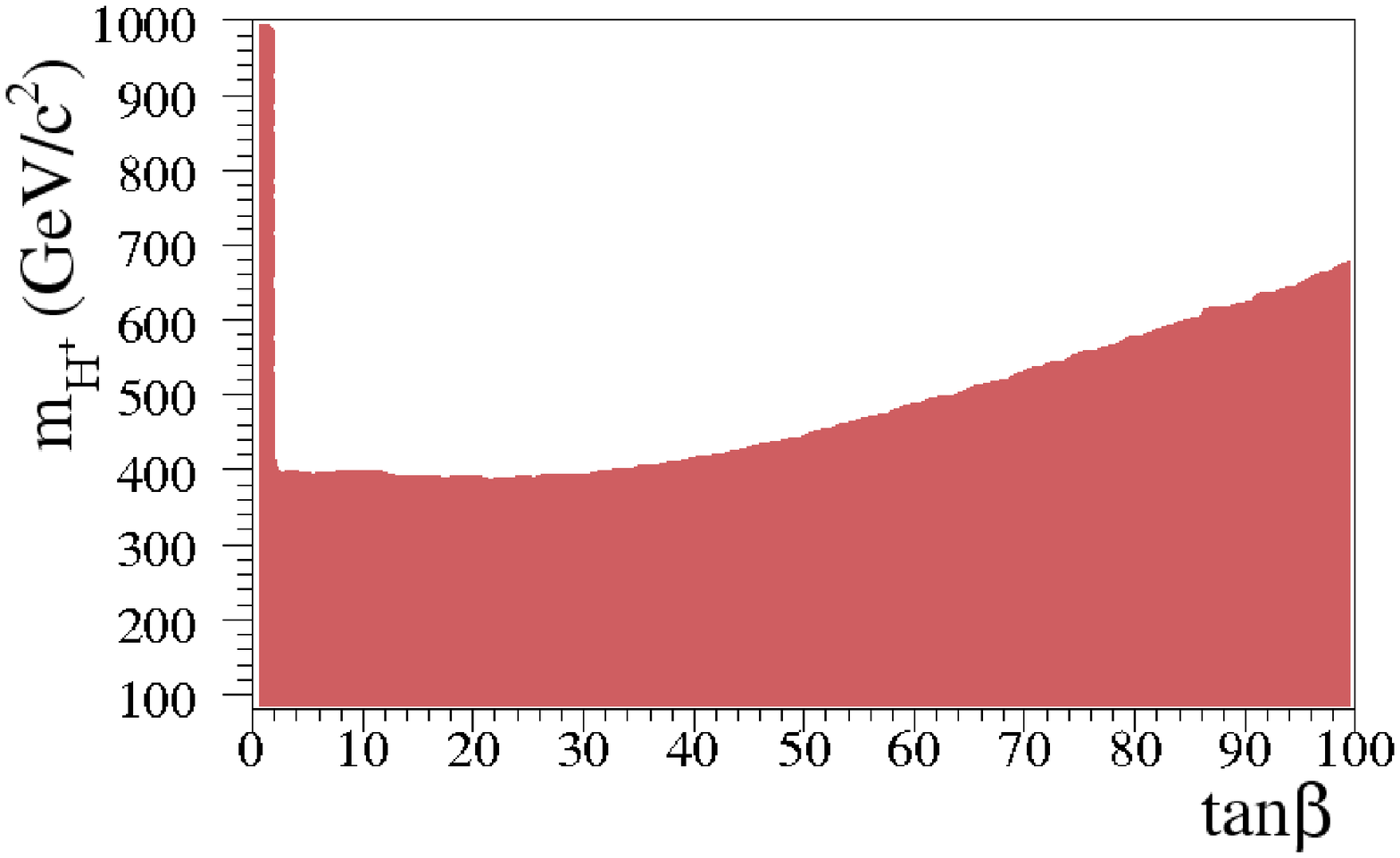}
\end{center}
\caption{Plots showing the excluded regions for the charged Higgs mass, as a function of $\tan\beta$, based on the 2HDM.  Left --- from the \babar\ measurement of $\BR(\Btn)$; Right --- Also including results from the study of $b\ra s\gamma$.} 

\label{fig:Exclusion}
\end{figure}

Beyond the Standard Model, the result can put a constraint on the mass of a charged Higgs boson, as a function of $\tan\beta$, as shown by equation \ref{eqn:2HDM}.  Figure \ref{fig:Exclusion} (left) shows the exclusion zone on the mass of the charged Higgs above the direct search limit from LEP, based on the results from \babar.  The exclusion can be further tightened by combining the results from $b\ra s\gamma$ with that from \Btn, and the combined exclusion is shown in figure \ref{fig:Exclusion} (right).

\section{The Decays $B\ra D^{(*)}\tau\nu$}

\subsection{Motivation}

Many semileptonic $B$ meson decays have been observed, but those involving a $\tau$ lepton remain experimentally challenging due to the presence of multiple neutrinos in the final state.  However due to measurements of semileptonic decays involving the lighter charged leptons, the branching fractions for $B\ra D^{(*)}\tau\nu$ can be predicted with high precision, and some of these predictions are shown in table \ref{tab:BDtn}.

\begin{table}
\caption{Standard Model branching fraction predictions for $\BDtn$.  Taken from~\protect\cite{chengeng}~\protect\cite{falk94}.} 
\label{tab:BDtn}
\begin{center}
\begin{tabular}{lr} \hline\hline
Decay Mode& $\BR$ (\%) \\ \hline
\Bzb$\to\Dm\taum\nutb$        & $0.69\pm 0.04$  \\
\Bzb$\to\Dstarm\taum\nutb$    & $1.41\pm 0.07$  \\
\Bm$\to\Dz\taum\nutb$         & $0.64\pm 0.04$  \\
\Bm$\to\Dstarz\taum\nutb$     & $1.32\pm 0.07$  \\ \hline\hline
\end{tabular}
\end{center}
\end{table}

\begin{figure}[h]
\begin{center}
\includegraphics[width=0.45\linewidth]{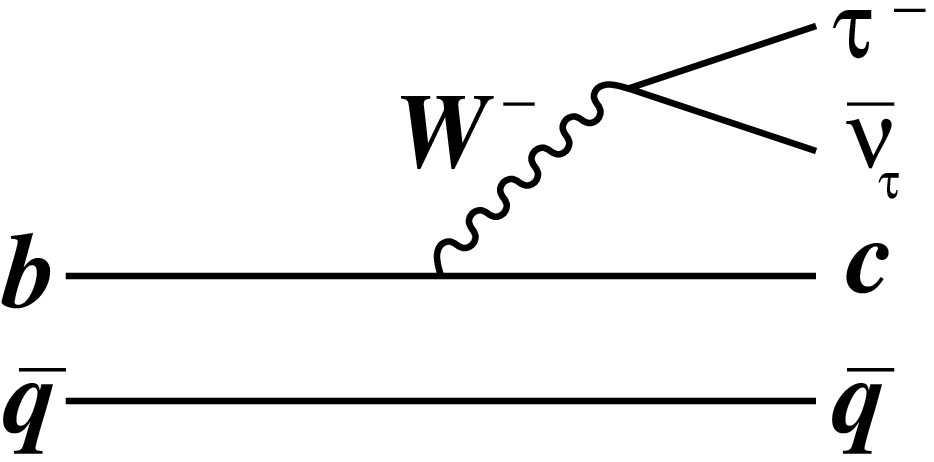}
\includegraphics[width=0.45\linewidth]{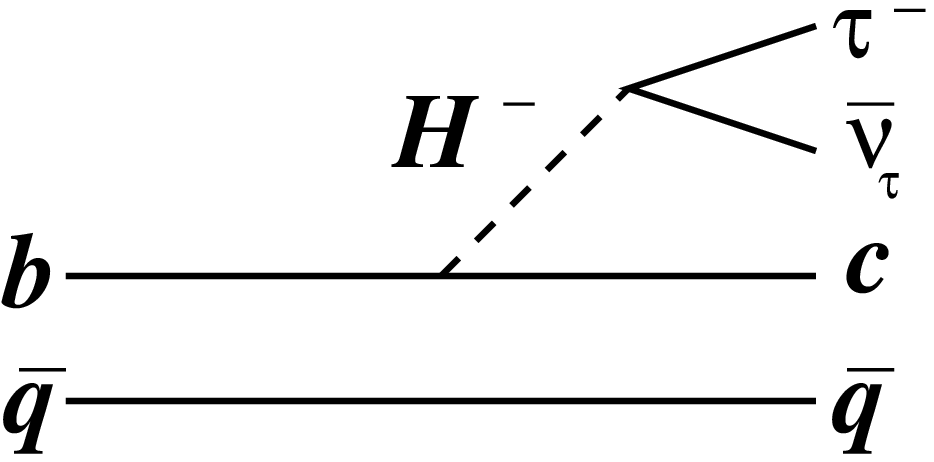}
\end{center}
\caption{Feynman diagrams showing the decay $\BDtn$:  left --- Standard Model diagram; right --- extra diagram featuring a beyond the Standard Model charged Higgs boson.}
\label{fig:BDtnFeyn}
\end{figure}

\begin{figure}[h]
\begin{center}
\includegraphics[width=0.6\linewidth]{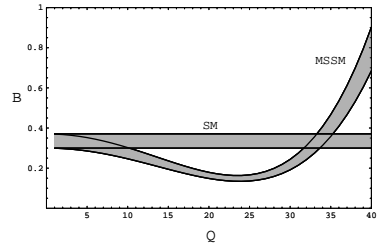}
\end{center}
\vspace{-0.5cm}
\caption{The ratio of $\BR(B\ra D\tau\nu) / \BR(B\ra D\ell\nu)$ in the Standard Model (SM), and in the Minimal SuperSymmetric Model (MSSM) including a charged Higgs boson, as a function of $Q=(m_{W}/m_{H}) \tan\beta$  \protect\cite{miura}.}
\label{fig:BDtnRatio}
\end{figure}

Beyond the Standard Model it is possible to introduce an extra diagram into the decay process by substituting the $W$ boson with a charged Higgs boson, shown in figure \ref{fig:BDtnFeyn}.  As the Higgs couples to mass, the effect of this diagram could be significant in semileptonic decays involving the $\tau$ lepton, even where no significant effect is seen in the decays involving electrons and muons.  Figure \ref{fig:BDtnRatio} shows the ratio of $\BR(B\ra D\tau\nu) / \BR(B\ra D\ell\nu)$, where $\ell=e,\mu$.

\subsection{Current Experimental Status}

The \babar\ analysis of \BDtn uses $232\times10^{6}$ \BB\ pairs, and utilises the hadronic set of tags described in section \ref{sec:recoil}, and can in addition make use of flavour correlation between $B_{reco}$ and the $D^{(*)}$ on the signal side.  A total of four decay channels are reconstructed: $B\ra D^{*+}\tau\nu$, $B\ra D^{*0}\tau\nu$, $B\ra D^{+}\tau\nu$, and $B\ra D^{0}\tau\nu$.  Only the leptonic decays of the $\tau$, i.e. $\tau\ra e\nu\nu$, and $\tau\ra\mu\nu\nu$ are used.
A simultaneous fit is carried out to all of these channels, and the branching fractions are normalised with respect to $D^{(*)}\ell\nu$ channels to eliminate certain systematic errors.

\begin{figure}[h]
\begin{center}
\includegraphics[width=0.6\linewidth]{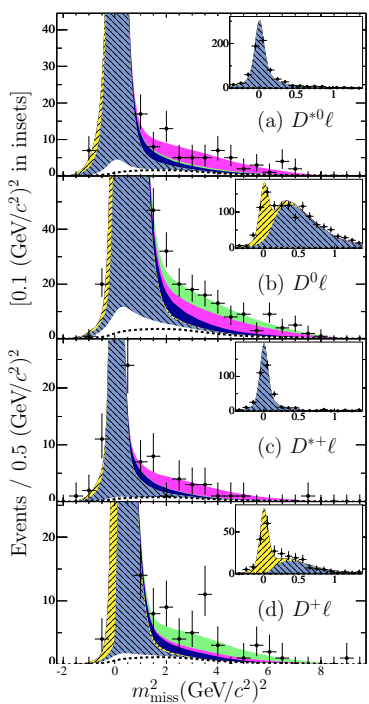}
\end{center}
\vspace{-0.5cm}
\caption{Distributions of events and fit projections in $m^{2}_{miss}$ 
for the four final states:
$\Dstarz\ell^-$, $\Dz\ell^-$, $\Dstarp\ell^-$, and $\Dp\ell^-$. 
The normalization region $m_{miss}\approx 0$ is shown with finer binning in the insets. 
The fit components are combinatorial background (white, below dashed line), 
charge crossfeed background (white, above dashed line), the $B\to D\ellm\nulb$ 
normalization mode (// hatching, yellow), the $B\to\Dstar\ellm\nulb$ normalization
mode ($\backslash\backslash$ hatching, light blue), $B\to D^{**}\ellm\nulb$ background (dark, or blue),
the $B\to D\taum\nutb$ signal (light grey, green), and the $B\to\Dstar\taum\nutb$ signal
(medium grey, magenta).  Plot taken from \protect\cite{babarBDtn}.} 
\label{fig:babarBDtn}
\end{figure}

The most discriminating variable is $m_{miss}$, the missing mass.  Figure \ref{fig:babarBDtn} shows the missing mass distributions for each of the four channels.

The results averaged over charged and neutral modes are: $\BR(B\ra D\tau\nu) = (0.86\pm0.24\pm0.11\pm0.06)\%$, and $\BR(B\ra D^{*}\tau\nu) = (1.62\pm0.31\pm0.10\pm0.05)\%$, corresponding to significances of 3.6$\sigma$ and 6.2$\sigma$ respectively.

\begin{figure}[h]
\begin{center}
\includegraphics[width=0.8\linewidth]{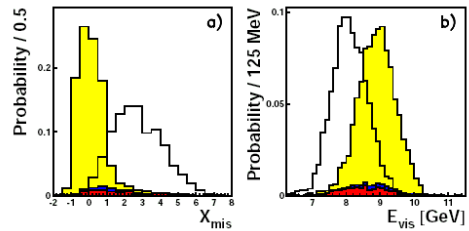}
\end{center}
\vspace{-0.5cm}
\caption{Signal (open histogram) and background (filled histograms) distributions from the Belle analysis of $\BDtn$ \protect\cite{belleBDtn}.} 
\label{fig:belleBDtn}
\end{figure}

The Belle analysis\cite{belleBDtn} looks for the decay $\Bzb\ra\Dstarp\taum\nutb$, using a sample of $535\times10^{6}$ \BB\ pairs.  The $\tau$ leptons are reconstructed in the decay modes $\tau\ra e\nu\nu$, and $\tau\ra\pi\nu$.  The most discriminating variables are $X_{mis}$, which is closely related to $m_{miss}$, and $E_{vis}$, the visible energy; the signal and background distributions in these variables are shown in figure \ref{fig:belleBDtn}.
The result obtained is 
$\BR(\Bzb\ra\Dstarp\taum\nutb) = (2.02^{+0.40}_{-0.37}\pm0.37)\%$, representing a significance of 5.2$\sigma$.

Current measurements of $\BDtn$ do not yet place any stong contraints on the possible presence of a charged Higgs boson, or its properties should it exist.

\section{Summary}
Leptonic and semileptonic decays of $B$ mesons into states involving $\tau$ leptons remain experimentally challenging, but can prove a useful tool for constraining Standard Model parameters, and also offer to constrain the effects of any new physics that may exist including the presence of a charged Higgs boson.  The current state of measurements of these decays is as follows:

\noindent \babar: 
\begin{eqnarray*}
\BR(\Btn) & = &  (1.2\pm0.4(\mathrm{stat.})\pm0.3(\mathrm{bkg.}) \\ & & \pm0.2(\mathrm{syst.}))\times10^{-4}, \\
\BR(B\ra D\tau\nu) & = & (0.86\pm0.24\pm0.11\pm0.06)\%, \\
\BR(B\ra D^{*}\tau\nu) & = & (1.62\pm0.31\pm0.10\pm0.05)\%.
\end{eqnarray*}
\noindent Belle:
\begin{eqnarray*}
\BR(\Btn) & = & (1.79^{+0.56}_{-0.49}(\mathrm{stat.}^{+0.46}_{-0.51}(\mathrm{syst.})) \\ & & \times10^{-4}, \\
\BR(\Bzb\ra\Dstarp\taum\nutb) & = & (2.02^{+0.40}_{-0.37}\pm0.37)\%.
\end{eqnarray*}


\bigskip 

\end{document}